# Combined Spectroscopic and Photometric Analysis of Flares in the Dwarf M Star EV Lacertae

**David Boyd**
*West Challow Observatory, OX12 9TX, UK; davidboyd@orion.me.uk*

**Robert Buchheim**
*Lost Gold Observatory, Gold Canyon, Arizona; bob@rkbuchheim.org*

**Sean Curry**
*Yank Gulch Observatory, Talent, Oregon; sxcurry@gmail.com*

**Frank Parks**
*Tierrasanta Astrophysics Observatory, San Diego, California; fgparks@mac.com*

**Keith Shank**
*Carollton, Texax; kasism@verizon.net*

**Forrest Sims**
*Desert Celestial Observatory, Gilbert, Arizona,; forrest@simsaa.com*

**Gary Walker**
*Maria Mitchell Observatory, Nantucket, Massachusetts; bailyhill14@gmail.com*

**John Wetmore**
*AZ Desertskies Observatory, Gilbert, Arizona; john@azdesertskies.com*

**James Jackman**
*ASU School of Earth and Space Exploration, Tempe, Arizona; jamesjackman@asu.edu*

**Abstract** We report results of an observing campaign to study the dwarf M flare star EV Lacertae. Between October 2021 and January 2022 we obtained concurrent B band photometry and low resolution spectroscopy of EV Lac on 39 occasions during 10 of which we observed flares with amplitude greater than 0.1 magnitude. Spectra were calibrated in absolute flux using concurrent photometry and flare-only spectra obtained by subtracting mean quiescent spectra. We measured B band flare energies between Log E = 30.8 and 32.6 erg. In the brightest flares we measured temporal development of flare flux in H I and He I emission lines and in the adjacent continuum and found that flux in the continuum subsided more rapidly than in the emission lines. Although our time resolution was limited, in our brightest flare we saw flux in the continuum clearly peaking before flux in the emission lines. We observed a progressive decrease in flare energy from Hβ to Hδ. On average we found 37% of B band flare energy appeared in the Hβ to Hε emission lines with the remainder contributing to a rise in continuum flux. We measured black-body temperatures for the brightest flares between 10,500 ± 700 K and 19,500 ± 500 K and found a linear relationship between flare temperature and continuum flux at 4170 Å. Balmer lines in flare-only spectra were well fitted by Gaussian profiles with some evidence of additional short-lived blue-shifted emission at the flare peak.

## 1. Introduction

Stellar flares are explosive events that occur when magnetic reconnection in the corona accelerates charged particles down into the chromosphere, heating the plasma and releasing energy across the electromagnetic spectrum (Benz and Güdel 2010; Allred *et al.* 2015). Flare output at visual wavelengths has been modelled as a combination of a fast, short-lived rise in the continuum produced by hot black-body radiation and a slower rise and decay in Balmer emission (see Kowalski *et al.* 2013 for references). Flares occur more often in stars of later spectral type, becoming most frequent in young, rapidly rotating, magnetically active M dwarfs (see for example results from TESS in Günther *et al.* 2020 and NGTS in Jackman *et al.* 2021). As M dwarfs are the most common stars in the galaxy, they are also the most common hosts of exoplanetary systems. The space weather environment around these stars will have a profound effect on the habitability of their planets and this has stimulated an increasing level of interest in understanding the nature and frequency of stellar flares.

EV Lac is a well-known flare star with mass 0.350 ± 0.020 $M_\odot$, radius 0.353 ± 0.017 $R_\odot$, luminosity 0.0128 ± 0.0003 $L_\odot$, and effective temperature 3270 ± 80 K (Paudel *et al.* 2021). It has a rotation period of 4.378 days (Pettersen 1980), faster than the 5.78-day mean rotation period of M dwarfs in both the K2 and SDSS surveys (Popinchalk *et al.* 2021). Fast rotation contributes to development of a strong magnetic field. Its spectral type has been variously described as dM3.5e (Reid *et al.* 1995), M4.0V (Lépine *et al.* 2013), and M4.5e (Joy and Abt 1974). Several multi-wavelength campaigns to observe flares in EV Lac have been published (see for example Paudel *et al.* 2021 and references therein) but have had limited success in recording flares concurrently with photometry and spectroscopy.

## 2. Observing campaign

Here we report the results of a campaign by a group of well-equipped amateur observers located in the USA and Europe, in which Jackman participated as our professional advisor and mentor, to specifically address that deficit by obtaining and analysing concurrent photometry and spectroscopy of EV Lac. The campaign was coordinated through biweekly online meetings and is part of a larger coordinated program of observations covering several flare stars. Members of the group obtained photometric and/or spectroscopic observations using the resources listed in Table 1 whenever circumstances permitted. Equipment is located at the observer's home unless stated otherwise. Photometric observations were reported to databases managed by the AAVSO (Kafka 2022) and BAA (BAA Photometry Database 2022). A shared project Google Drive was used to manage spectroscopic observations, including a timeline recording when concurrent photometric and spectroscopic observations had been obtained.

Observations reported here run from October 2021 to January 2022. During that time, we recorded 107 photometry sessions and 72 spectroscopy sessions including 39 in which photometry and spectroscopy were obtained concurrently. In these 39 sessions we identified 10 containing flares with B-magnitude amplitudes greater than 0.1 magnitude and which form the basis of this analysis. A journal of these ten sessions is given in Table 2. Analysis of our data was performed with custom PYTHON software which made extensive use of the Astropy package (Astropy Collaboration 2018).

## 3. Photometric observations

Photometric observations were mostly made with 0.35- and 0.5-m telescopes, using Astrodon dielectric Johnson-Cousins (J-C) B-band photometric filters. This passband was chosen as light output from flares increases towards shorter wavelengths (Paudel *et al.* 2021; Kowalski *et al.* 2013) but recording efficiency in the UV passband is generally low with our observing equipment and with atmospheric transmission at our low altitudes. A small number of observations were made with smaller telescopes using J-C V band filters to observe changes in the color index of EV Lac during flares. Photometric images were bias, dark, and flat corrected and instrumental magnitudes obtained by aperture photometry using the software AIP4WIN (Berry and Burnell 2005) or MAXIM DL (Diffraction Limited 2023). Comparison star magnitudes were obtained from the AAVSO chart for EV Lac (AAVSO 2022) and used to convert instrumental to standard magnitudes in the J-C system. In order to establish a consistent timeframe between datasets recorded concurrently, observation times were obtained from internal computer clocks regularly synchronized to internet time servers (NIST, NPL 2023) and were recorded in FITS headers as Julian Date (JD). Heliocentric corrections were not applied. Exposures ranged between 20 and 120 seconds depending on aperture used and conditions. B band photometric observations listed in Table 2 and used in this analysis totalled 38.7 hr.

## 4. Spectroscopic observations

Spectroscopic observations covered the wavelength range 3750 Å to 7000 Å and were made with ALPY (R ~ 500) and LISA (R ~ 1000) spectroscopes (Shelyak Instruments 2022) auto-guided on 0.3- and 0.4-m telescopes using 23-μ slits to match typical atmospheric seeing at the observing sites. Spectra were usually integrated for 300 seconds. Spectra were processed with the ISIS spectral analysis software (Buil 2021). Spectroscopic images were bias, dark, and flat corrected, geometrically corrected, sky background subtracted, spectral profile extracted, and wavelength calibrated using integrated ArNe calibration sources. Spectra of a nearby star with a known spectral profile from the MILES library of stellar spectra (Falcón-Barroso *et al.* 2011) situated as close as possible in airmass to EV Lac at the time of observation were obtained both immediately before and immediately after the spectra of EV Lac. By adopting a parameterization of atmospheric transmission as a function of airmass (Vidal-Madjar 2010), we were able to correct for instrumental and atmospheric losses at the airmass of each spectral image. Spectroscopic observations listed in Table 2 and used in to this analysis totalled 40.8 hr.

Table 1. Equipment used by members of the group.

| *Observer* | *Photometry Equipment* | *Spectroscopy Equipment* |
|---|---|---|
| Boyd (DB) | 0.35 m SCT + B filter | 0.28 m SCT + LISA |
| Buchheim (RB) | — | 0.41 m SCT + ALPY |
| Curry (SC) | 0.11 m refractor + B, V filters | 0.11 m refractor + ALPY |
| Parks (FP) | 0.11 m refractor + B filter | 0.2 m Newtonian + LISA |
| Shank (KS) | — | 0.35 m SCT + LISA |
| Sims (FS) | 0.11 m refractor + B, V, $R_c$ filters | 0.35 m CDK + LISA |
| Walker (GW) | 0.5 m CDK + U, B, V, $R_c$ filters at Sierra Remote Observatory, Auberry, CA | — |
| Wetmore (JW) | — | 0.28 m SCT + LISA |

Table 2. Journal of photometric and spectroscopic observations used in this analysis.

| *Observing Session* | *Date* | *Start of Photometry (JD)* | *Duration of Photometry (hr)* | *No. of Images* | *Band* | *Observer Initials* | *Start of Spectroscopy (JD)* | *Duration of Spectroscopy (hr)* | *No. of Spectra* | *Resolving Power* | *Observer Initials* |
|---|---|---|---|---|---|---|---|---|---|---|---|
| 1 | 2021 Oct 30 | 2459517.675 | 3.925 | 360 | B | GW | 2459517.676 | 4.217 | 51 | 500 | RB |
| 2 | 2021 Nov 4 | 2459522.686 | 3.922 | 360 | B | GW | 2459522.655 | 4.385 | 53 | 500 | RB |
| 3 | 2021 Nov 13 | 2459531.666 | 3.904 | 360 | B | GW | 2459531.593 | 5.695 | 62 | 500 | RB |
| 4 | 2021 Nov 15 | 2459533.665 | 3.900 | 360 | B | GW | 2459533.623 | 4.559 | 55 | 500 | RB |
| 5 | 2021 Nov 18 | 2459536.665 | 3.917 | 360 | B | GW | 2459536.599 | 4.996 | 58 | 500 | RB |
| 6 | 2021 Nov 21 | 2459540.296 | 4.378 | 530 | B | DB | 2459540.301 | 4.180 | 50 | 1000 | DB |
| 7 | 2021 Nov 26 | 2459544.608 | 3.903 | 1405 | B & V | FP FS GW | 2459544.580 | 5.000 | 57 | 500 | RB |
| 8 | 2021 Dec 12 | 2459560.557 | 5.254 | 160 | B | GW | 2459560.636 | 2.423 | 24 | 1000 | JW |
| 9 | 2022 Jan 13 | 2459593.261 | 3.033 | 410 | B | DB | 2459593.262 | 3.159 | 38 | 1000 | DB |
| 10 | 2022 Jan 14 | 2459594.280 | 2.546 | 350 | B | DB | 2459594.287 | 2.043 | 25 | 1000 | DB |

## 5. Analysis of photometric data

Visual examination of the photometric light curves in the 39 sessions with concurrent spectroscopy identified 10 sessions in which flares rose above the quiescent level with B magnitude amplitudes greater than 0.1 magnitude. This threshold was chosen as our subsequent analysis found that, at the low resolving power of our spectra, poorly-defined or lower amplitude flares did not yield spectra of sufficient quality for the quantitative analysis described here. In these 10 sessions we identified the 12 flares shown in Figure 1. Flares come in many forms, ranging from rapidly rising and falling to slowly rising and gradually decaying, with new flares starting before quiescence is reached. The start and end times of flares were identified by visual inspection of the photometric light curves as the times at which the flux level started to rise above the quiescent level and either returned to the quiescent level, a second flare began, or the observing session finished. All light curves were thus divided into flares and quiescent regions. The regions identified as flares are marked in red in Figure 1. The magnitude scale of each light curve in Figure 1 is chosen to show maximum detail. All were recorded with similar sized telescopes so have similar noise levels.

The median quiescent B magnitude during each observing session was calculated and converted to an absolute quiescent B magnitude using the distance modulus of EV Lac determined from its distance of 5.05 parsec derived from the parallax measured by Gaia (Bailer-Jones *et al.* 2021). The mean B band quiescent luminosity in erg/s during each observing session was calculated from the absolute quiescent B magnitude using B band solar luminosity and absolute solar B magnitude on the Vegamag system (Bohlin and Gilliland 2004) as transmitted through the same B band filter profile used for our observations. The mean B band quiescent magnitude and luminosity over all observing sessions were $11.89 \pm 0.04$ mag and $3.14 \pm 0.11 \times 10^{29}$ erg/s, respectively. The small uncertainties indicate that the quiescent energy output of EV Lac was relatively stable between October 2021 and January 2022.

Each photometric B magnitude was converted to a B band luminosity in the same way and the B band luminosity of any flare present was obtained by subtracting the mean B band quiescent luminosity for that session. These B band flare luminosities were integrated over the time span of each photometric exposure to find the energy in erg contributed to the flare by that exposure. The total energy emitted by the flare in the B band was then found by integrating these contributions through the duration of the flare. Table 3 gives information about times, magnitudes, and energies of the 12 flares. It also includes measurements of the $t_{1/2}$ and equivalent duration parameters which are described in section 8.

We also recorded a series of V magnitude measurements concurrently with B magnitudes on 26 November 2021, enabling us to derive the B–V color index shown in Figure 2. The uncertainty on individual B–V values was 0.02 mag. The mean B–V color index of EV Lac during quiescence prior to the first flare on that date was $1.66 \pm 0.02$ mag. Given the consistency of our quiescent B magnitudes noted in Table 3, we assume this to be representative of the quiescent color index of EV Lac on other dates. In Figure 2 we show B–V peaking at 1.37 mag and 0.98 mag during the two flares recorded on that date.

## 6. Analysis of spectroscopic data

The mean B magnitude during each spectrum was calculated by converting photometric B magnitudes obtained within the exposure time of the spectrum to fluxes, averaging these fluxes over the duration of the spectrum, and converting this back to a B magnitude. Using the procedure described in Boyd (2020), each spectrum was then calibrated in absolute flux in FLAM units as erg/cm$^2$/s/Å using this concurrently obtained mean B magnitude. This procedure made use of CALSPEC spectra (Bohlin *et al.* 2014) to establish a zero point B magnitude for the B band filter used for these observations. Given the relatively small distance of EV Lac we assume negligible interstellar reddening. Reiners *et al.* (2018) report a radial velocity for EV Lac of 0.19 km/s, and the heliocentric radial velocity of EV Lac varied by less than 10 km/s during our observations. As these are below a level which would affect our analysis, no velocity corrections were made.

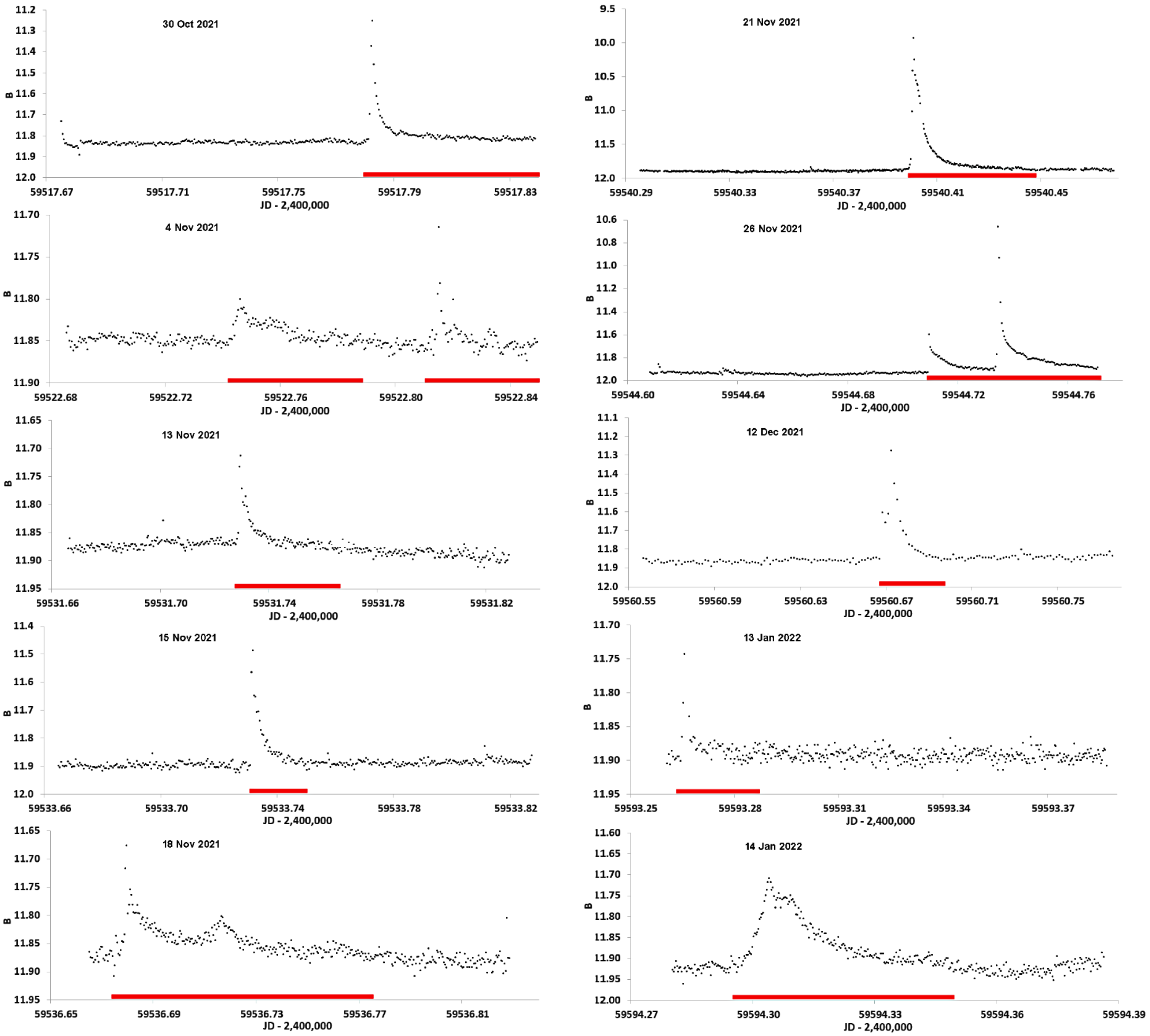

Figure 1. B magnitude light curves of 12 EV Lac flares showing in red the regions identified as flares.

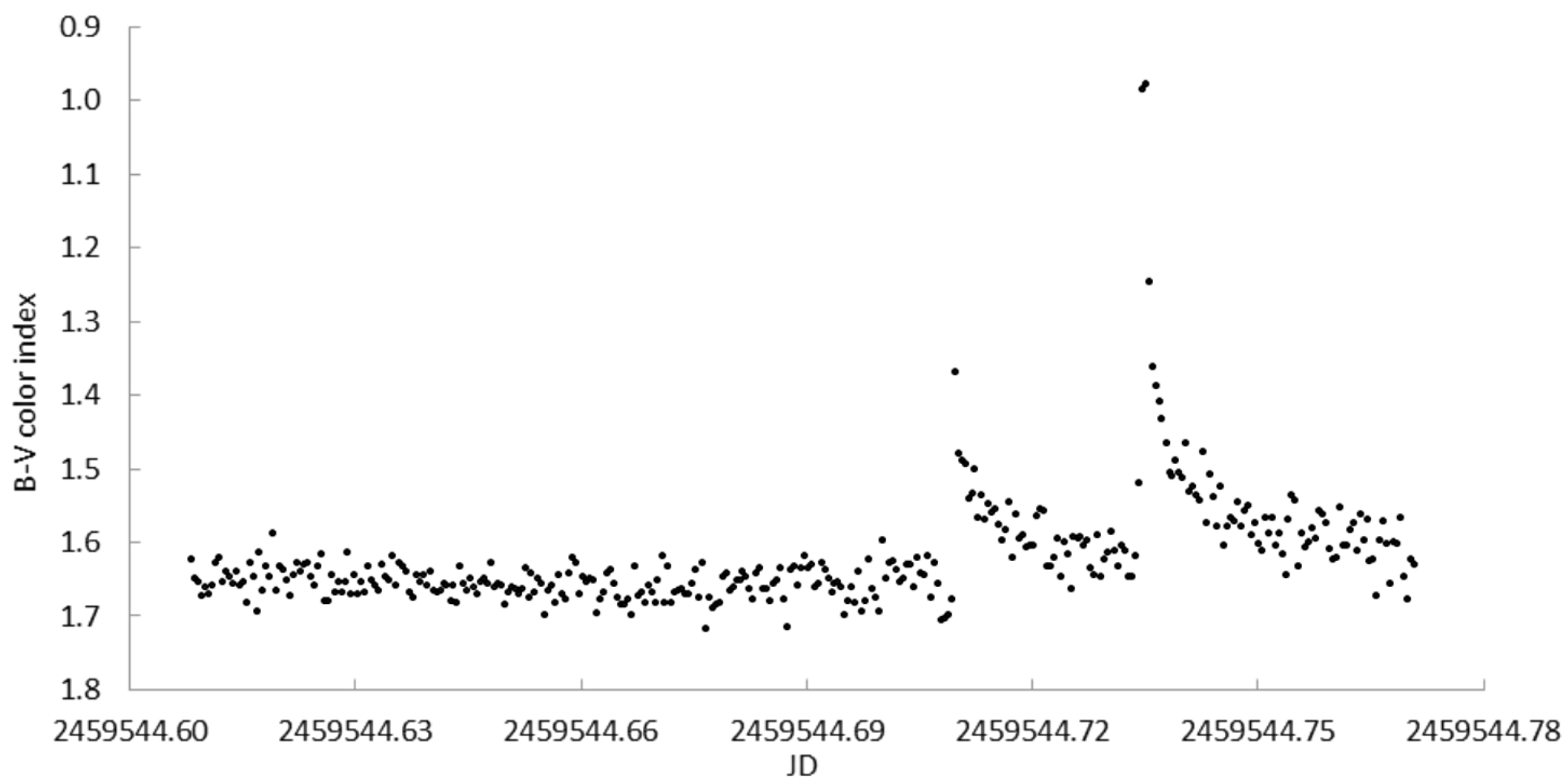

Figure 2. B–V color index of EV Lac on 26 November 2021.

Table 3. Parameters of 12 recorded flares of EV Lac.

| *Flare No.* | *Date* | *Start Time of Flare (JD)* | *Rise Time of Flare (min)* | *Decay Time of Flare (min)* | *Quiescent B-band Mag. (mag)* | *Peak B-band Mag. (mag)* | *B-band Magnitude (mag)* | *B-band Amplitude Log (erg)* | *B-band $t_{1/2}$ (min)* | *Equivalent Duration* |
|---|---|---|---|---|---|---|---|---|---|---|
| 1 | 2021 Oct 30 | 2459517.779 | 4.6 | 80.9 | 11.84 | 11.25 | 0.58 | 31.90 | 1.95 | 4.0 |
| 2 | 2021 Nov 4 | 2459522.741 | 7.2 | 57.7 | 11.85 | 11.80 | 0.05 | 31.17 | 21.03 | 0.8 |
| 3 | 2021 Nov 4 | 2459522.811 | 5.9 | 34.1 | 11.85 | 11.72 | 0.14 | 30.77 | 1.20 | 0.3 |
| 4 | 2021 Nov 13 | 2459531.727 | 3.3 | 48.9 | 11.88 | 11.71 | 0.17 | 31.33 | 3.82 | 1.1 |
| 5 | 2021 Nov 15 | 2459533.73 | 1.9 | 28.0 | 11.89 | 11.49 | 0.40 | 31.55 | 2.43 | 1.9 |
| 6 | 2021 Nov 18 | 2459536.675 | 7.2 | 139.4 | 11.87 | 11.68 | 0.20 | 31.94 | 3.82 | 4.6 |
| 7 | 2021 Nov 21 | 2459540.397 | 5.4 | 68.8 | 11.90 | 9.93 | 1.97 | 32.61 | 1.67 | 22.0 |
| 8 | 2021 Nov 26 | 2459544.709 | 0.7 | 30.1 | 11.94 | 11.60 | 0.34 | 31.72 | 2.44 | 2.9 |
| 9 | 2021 Nov 26 | 2459544.733 | 2.0 | 52.2 | 11.94 | 10.66 | 1.28 | 32.23 | 1.34 | 9.5 |
| 10 | 2021 Dec 12 | 2459560.667 | 7.9 | 39.6 | 11.86 | 11.28 | 0.58 | 32.07 | 5.31 | 6.1 |
| 11 | 2022 Jan 13 | 2459593.264 | 2.9 | 27.1 | 11.90 | 11.69 | 0.20 | 31.05 | 1.99 | 0.6 |
| 12 | 2022 Jan 14 | 2459594.296 | 11.4 | 70.1 | 11.92 | 11.71 | 0.22 | 31.94 | 15.84 | 4.8 |

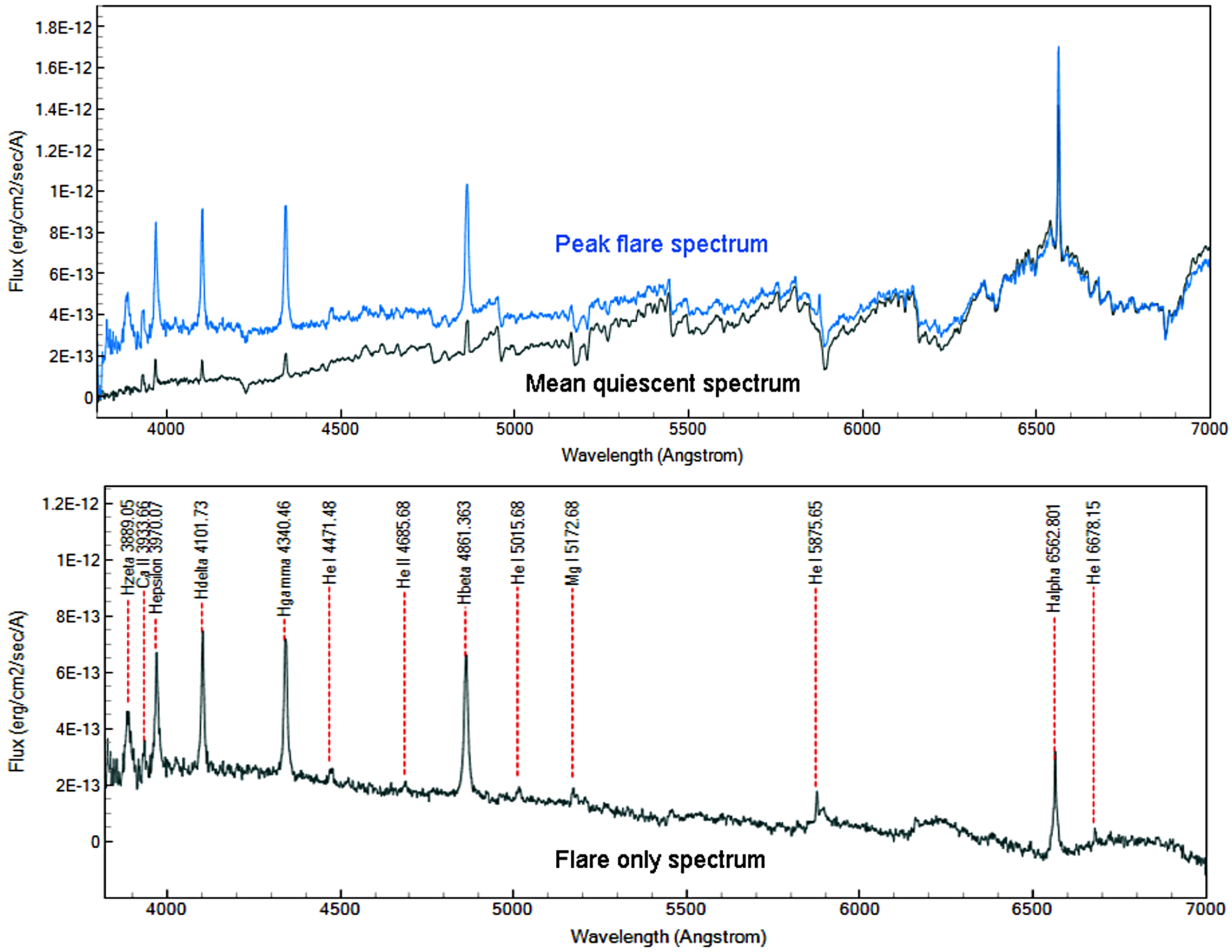

Figure 3. Mean absolute flux quiescent and peak flare spectra of EV Lac on 21 November 2021 (upper) and the peak flare-only spectrum with identified emission lines (lower).

For each observing session, all our absolute flux spectra during quiescence were averaged to find a mean absolute flux quiescent spectrum. Given that all spectra in a session are likely to have been recorded under similar conditions and processed taking account of varying airmass, we consider the standard deviation of quiescent flux at each wavelength to give a realistic estimate of the uncertainty in measuring the flux at that wavelength for all spectra in that session. By averaging over the wavelength range of each Balmer line we could also obtain an estimate of the uncertainty in the flux in these lines. Dividing the mean absolute quiescent flux in each observing session by its standard deviation at each wavelength gives an estimate of the SNR of spectral flux at that wavelength for that session. We found SNR to vary between 10 and 30 for most sessions except below ~4000 Å, where throughput started to fall due to declining equipment efficiency.

## 7. Calculating flare-only spectra

The mean absolute flux quiescent spectrum of EV Lac for the observing session on 21 November 2021 is shown in Figure 3 (upper). TiO molecules form in the atmosphere of cool M-type stars and produce the deep absorption bands seen in this spectrum (Gray and Corbally 2009). Also shown is the spectrum recorded at the peak of the flare. Subtracting the mean quiescent spectrum from the peak flare spectrum gives the peak flare-only spectrum.

This is shown in Figure 3 (lower) which identifies H I and He I emission lines plus a weak line of He II 4686 Å and possibly the Mg I triplet at 5167, 5173, and 5184 Å (Gray and Corbally 2009). See Table 5 below for the energy emitted in lines which could be measured reliably. The likely presence of He II in emission indicates a high temperature. The "humps" in the flare-only spectrum above 6000 Å are likely to be the result of TiO absorption bands becoming shallower during a flare relative to their depth in quiescence because of molecular dissociation during the flare.

As a check on our measurements of B-band flare energy from photometry in section 5, each flare-only spectrum was multiplied by the transmission profile of our B filter to give the B band flux in the spectrum in $erg/cm^2/s$. This was integrated over the time interval of each spectrum and multiplied by $4\pi d^2$, where d is the distance to EV Lac, to give the B band energy in each flare-only spectrum in erg. This assumes energy is being emitted isotropically into a sphere of radius d, although in practice emission from the flare is likely to be anisotropic. Nevertheless, it is conventional to assume isotropic emission for the purpose of calculating total energy emission. Integrating the energy recorded in each flare-only spectrum over all spectra in the flare gives a consistency check on the total B band energy in the flare. Comparing this with the measurement we obtained for the B band flare energy from photometry we find that, averaging over all flares, the two estimates of flare energy agree to within 2%.

## 8. Empirical flare parameters

Several parameters have been proposed in the literature to characterize properties of flares. One is $t_{1/2}$, defined by Kowalski *et al.* (2013) as the time interval between half maximum on the rise of the flare and the same height on its decay, in other words, the duration of the flare measured at half maximum. This is independent of the shape of the flare profile. We measured the $t_{1/2}$ times for the flares in our B band photometry and these are listed in Table 3.

Another measure that has been widely adopted for the longevity of flares is the equivalent duration defined in Gershberg (1972) as the ratio of flare-only energy in a specific band, in our case the B band, to quiescent luminosity in the same band. This is also independent of the flare profile. Table 3 contains our measurements of equivalent duration for each flare.

Table 4. Black-body temperatures for the four most energetic peak flare-only spectra.

| *Flare No.* | *Date* | *JD of Spectrum* | *Black-body Temperature (K)* |
|---|---|---|---|
| 7 | 2021 Nov 21 | 2459540.399 | 19,500 ± 500 |
| 7 | 2021 Nov 21 | 2459540.403 | 13,300 ± 600 |
| 9 | 2021 Nov 26 | 2459544.738 | 12,300 ± 400 |
| 10 | 2021 Dec 12 | 2459560.674 | 10,500 ± 700 |

## 9. Black-body temperature of continuum during flares

To estimate the equivalent black-body temperature of the flare-only continuum during a flare, we performed a non-linear least-squares fit of a Planck function to the continuum of flare-only spectra in the region 4120–5150 Å, excluding any emission or absorption features. In most cases the flux level of the individual flare-only spectra was too low to yield a reliable fit. However, we were able to obtain reasonable fits of black-body temperatures for four spectra at the peak of the three most energetic flares numbered 7, 9, and 10 in Table 3. These temperatures are listed in Table 4 and Figure 4 shows the black-body spectrum fitted to the peak flare-only spectrum on 21 November 2021. The uncertainty in temperature is from the covariance in the non-linear least squares fit. In Table 4, the first spectrum of flare 7 is at the flare peak, while the second immediately follows the peak.

We also attempted to fit a Planck function to a similar region of the continuum for each of the mean quiescent spectra, excluding emission or absorption features. The mean black-body temperature and standard deviation we found over all quiescent spectra was 3097 ± 251 K. From Pecaut and Mamajek (2013) the mean quiescent B–V color index of 1.66 mag observed on 26 November 2021 corresponds to spectral type M4V and effective temperature around 3200 K. Given the difficulty of measuring the low flux levels in this region of quiescent spectra, we consider the agreement with spectral type M3.5V and effective temperature 3270 ± 80 K given in Paudel *et al.* (2021) to be acceptable.

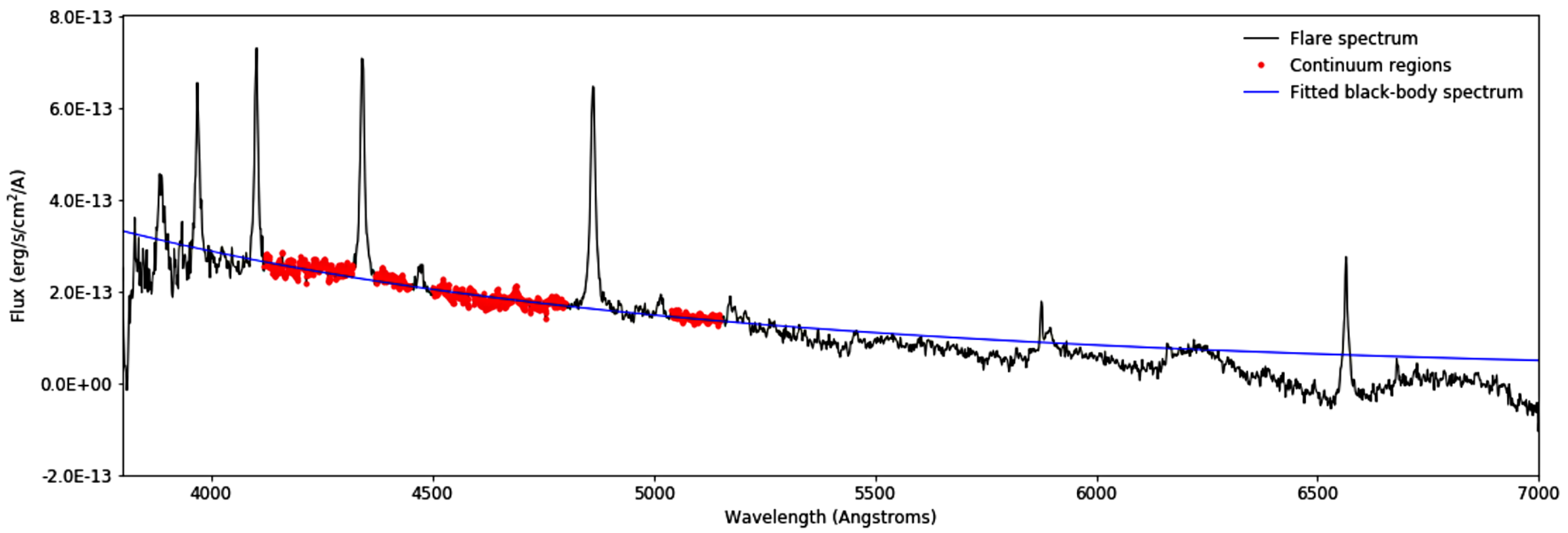

Figure 4. Fitted black-body spectrum for the peak flare-only spectrum on 21 November 2021 showing the continuum regions used for the fit.

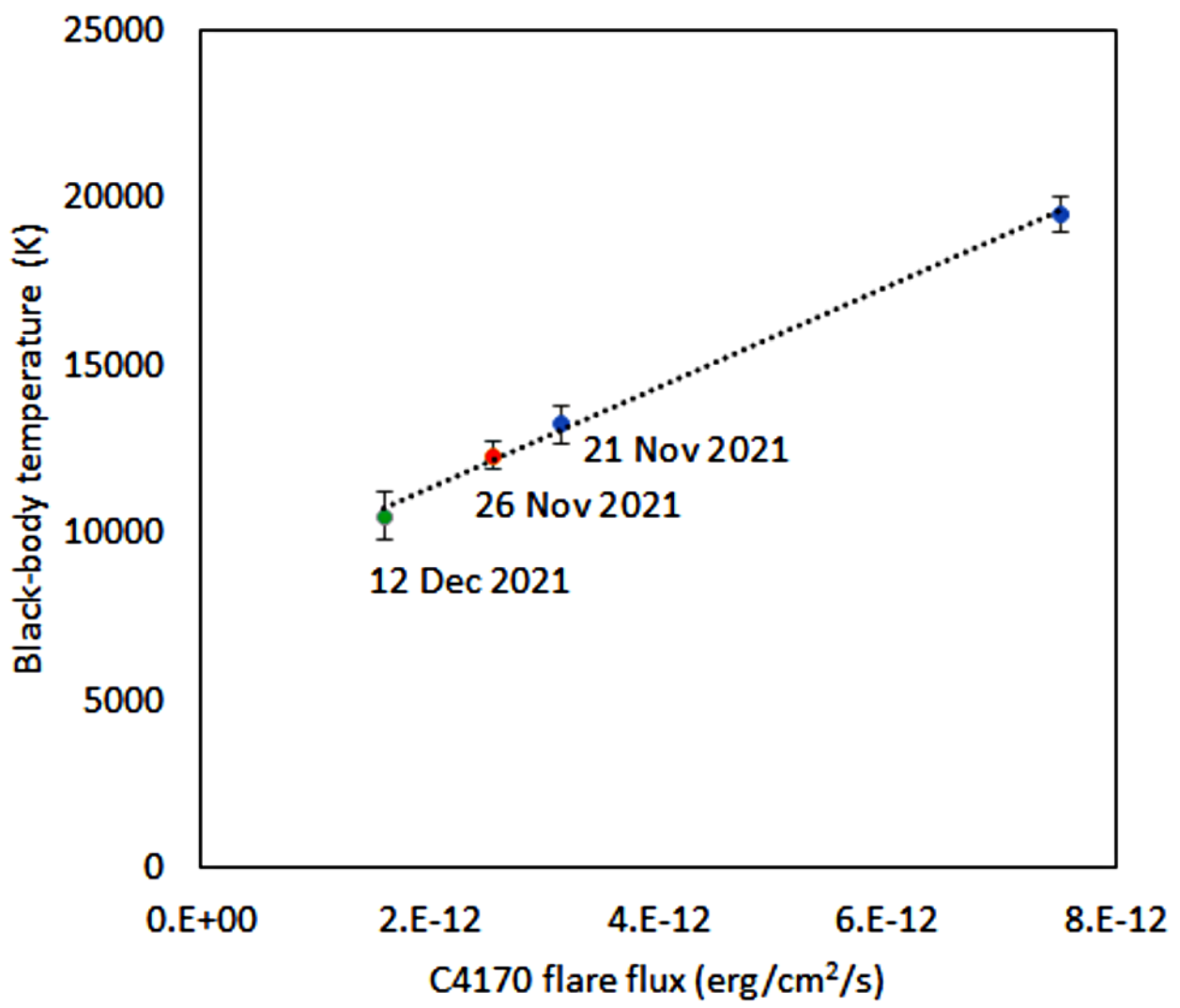

Figure 5. Black-body temperature vs flare flux in the C4170 region for the four most energetic flare-only spectra.

The peak black-body temperature of 12,300 K on 26 November 2021 contrasts with an effective temperature of around 4850 K from the peak B–V color index of 0.98 (Pecaut and Mamajek 2013). Whereas the peak black-body temperature is derived from the spectral energy continuum profile at the peak of the flare, the B-V color index is an indication of the effective temperature of the M dwarf star as a whole, increased above its quiescent level by the presence of the flare.

Kowalski *et al.* (2013) defined a region of the blue continuum labeled C4170 centered on 4170 Å with width 30 Å which could be used to provide a measure of flux level in the continuum. We integrated the flux in this region under the four flare-only spectra in Table 4 and used this to investigate a potential correlation between flare continuum flux in this region and black-body temperature at the peak of a flare. Figure 5 shows that, for flares in this temperature range, there does appear to be a linear relation between the black-body temperature derived from a fit to the continuum and the integrated flux in the C4170 region of the continuum.

In each of these three flares the black-body temperature of the following spectrum recorded five minutes later had dropped to below 4000 K and the integrated flux in the C4170 region had fallen below $10^{-12}$ erg/cm$^2$/s. This demonstrates how quickly temperature in a flare drops and energy in the flare dissipates after the initial sharp release of energy.

## 10. Analysis of flare energy in emission lines

In previous studies, higher resolving powers have often been used to examine in detail the behavior of individual emission lines (see for example Johnson *et al.* 2021). Working at lower resolving power and covering a wide wavelength range, we record several Balmer lines in our spectra. To find the energy emitted during a flare in a specific emission line, we first linearly interpolated the continuum under the line between regions of the continuum outside the line and integrated the area between the line profile and the interpolated continuum to obtain the integrated flux in the line in erg/cm$^2$/s. In doing this we were careful to set the continuum regions used for interpolation far enough away from the peak wavelength of the line that they did not include wings of the line which expanded at the peak of a flare, as shown in Figure 10. We then did the same with the mean quiescent spectrum to find the integrated flux in the line during quiescence and subtracted this from the integrated flux in the line to obtain the flux in the line from the flare in erg/cm$^2$/s. The flare flux in the line was then multiplied by the time interval between spectra and integrated over all spectra in the flare to get the total flux emitted by the flare in the line in erg/cm$^2$. Finally, this was multiplied by $4\pi d^2$, where d is the distance to EV Lac, to give the total energy in erg emitted by the flare in that emission line, again assuming isotropic emission.

The uncertainty in measuring flare energy in emission lines accrues mainly from two sources. One is the uncertainty in the flux level at each line as determined from the standard deviation in quiescent flux described earlier. The other is the uncertainty in defining the level of the interpolated continuum under emission lines because of local variations in the continuum on either side of the line. Both these sources propagate into the uncertainty in flare energy in an emission line.

Table 5 lists flare energy in the Hα to Hε Balmer lines for each flare where this is measurable. At our resolving power, the Hε line is blended with the Ca II H line, while the nearby Ca II K line is well resolved. On the basis that the two calcium lines have broadly similar strength (Rauscher and Marcy 2006), we constructed a pseudo Hε line, labelled ~Hε, by subtracting the Ca II K flux from the Hε + Ca II H line flux. There is visible evidence in some of the R = 1000 spectra of emission lines of He I 4471, 5016, 5876, and 6678 Å and He II 4686 Å, but only the He I 5876 Å line yields credible values in some of the larger flares and these are also included in Table 5. Figure 6 shows that, particularly in the more energetic flares, flare energy decreases progressively from Hβ to Hδ.

Table 5. Energy emitted in H I and He I 5876 Å emission lines during each flare.

| *Flare No.* | *Date* | *Hα (×10$^{30}$ erg)* | *Hβ (×10$^{30}$ erg)* | *Hγ (×10$^{30}$ erg)* | *Hδ (×10$^{30}$ erg)* | *~Hε (×10$^{30}$ erg)* | *He I 5876 (×10$^{30}$ erg)* |
|---|---|---|---|---|---|---|---|
| 1 | 2021 Oct 30 | 9.3 ± 1.8 | 8.2 ± 0.4 | 6.8 ± 0.4 | 6.4 ± 0.4 | 5.5 ± 1.1 | — |
| 2 | 2021 Nov 4 | 1.1 ± 1.9 | 4.4 ± 0.5 | 2.6 ± 0.4 | 2.5 ± 0.4 | 1.9 ± 1.2 | — |
| 3 | 2021 Nov 4 | — | 0.6 ± 0.3 | 0.1 ± 0.4 | 0.4 ± 0.4 | — | — |
| 4 | 2021 Nov 13 | 9.8 ± 2.8 | 5.2 ± 0.6 | 2.4 ± 0.8 | 1.8 ± 0.7 | 0.1 ± 2.2 | — |
| 5 | 2021 Nov 15 | 5.6 ± 1.3 | 4.3 ± 0.3 | 3.9 ± 0.3 | 2.3 ± 0.3 | 2.4 ± 0.8 | — |
| 6 | 2021 Nov 18 | 5.5 ± 2.7 | 12.6 ± 0.7 | 10.6 ± 0.6 | 8.0 ± 0.6 | 7.3 ± 1.6 | — |
| 7 | 2021 Nov 21 | 45.7 ± 1.2 | 47.0 ± 0.9 | 36.9 ± 1.4 | 25.1 ± 1.5 | 18.1 ± 11.6 | 15.4 ± 0.1 |
| 8 | 2021 Nov 26 | 7.9 ± 0.8 | 5.9 ± 0.3 | 4.8 ± 0.4 | 3.5 ± 0.5 | 0.8 ± 0.9 | 1.9 ± 0.1 |
| 9 | 2021 Nov 26 | 32.9 ± 1.3 | 26.8 ± 0.6 | 18.3 ± 0.7 | 13.3 ± 0.9 | 4.5 ± 1.6 | 4.3 ± 0.1 |
| 10 | 2021 Dec 12 | 8.8 ± 1.0 | 11.0 ± 0.6 | 8.6 ± 0.8 | 5.7 ± 0.8 | 6.2 ± 2.1 | 5.8 ± 0.1 |
| 11 | 2022 Jan 13 | 0.7 ± 1.0 | 1.6 ± 0.5 | 1.3 ± 0.5 | 1.5 ± 0.6 | 1.5 ± 4.3 | — |
| 12 | 2022 Jan 14 | 28.0 ± 3.7 | 18.6 ± 1.0 | 13.5 ± 0.9 | 12.5 ± 1.3 | 9.2 ± 8.2 | 2.8 ± 0.2 |

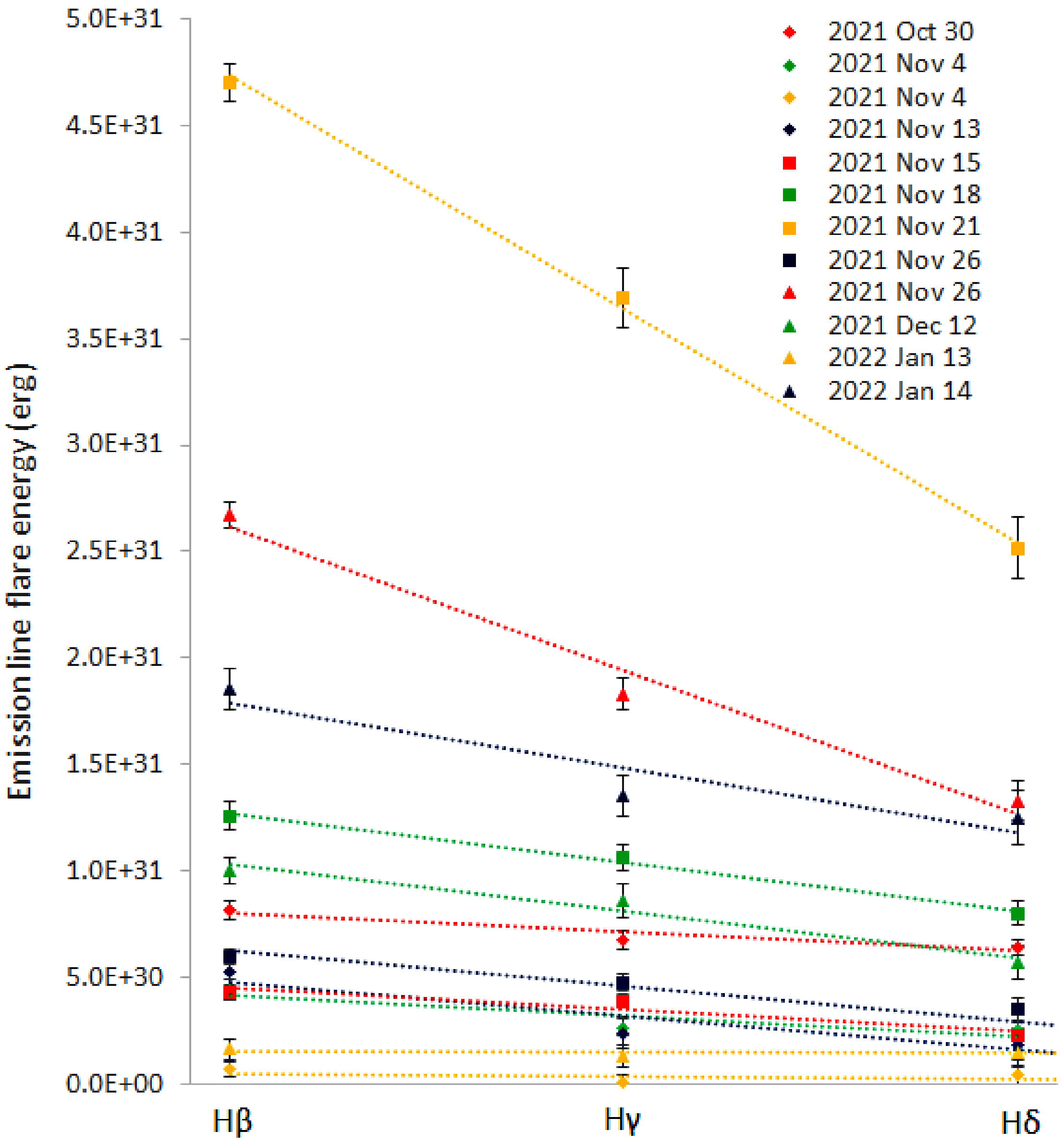

Figure 6. Decrement of Balmer line flare energy from Hβ to Hδ.

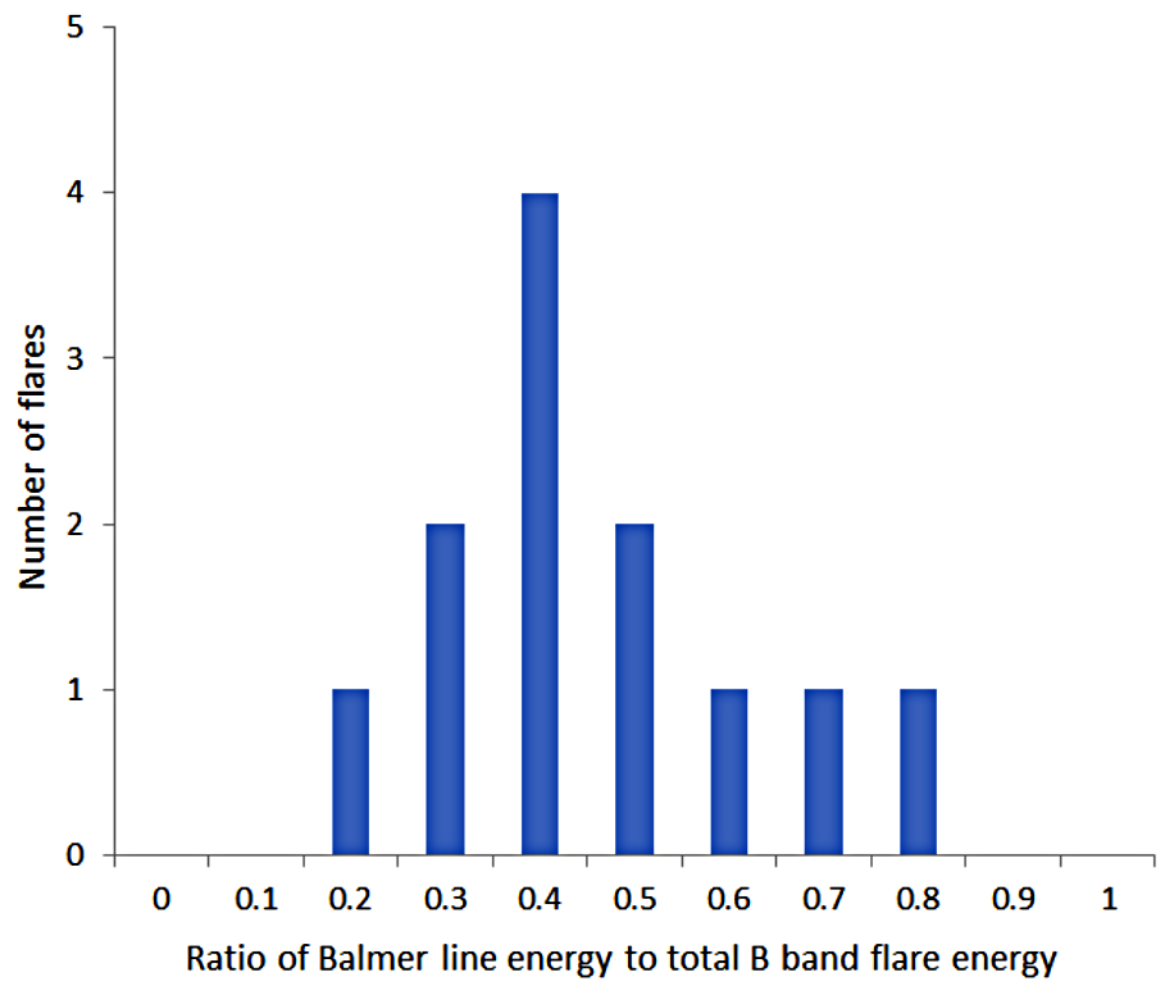

Figure 7. Histogram of the ratio for each flare of the total flare energy in the Hβ to Hε lines to the total flare energy emitted in the B band.

For each flare we aggregated the total flare energy in the Hβ to Hε lines, all of which lie within the B band, and calculated the ratio of this to the total flare energy emitted in the B band. Figure 7 shows a histogram of this ratio for all flares. The median percentage contribution of these emission lines to the total energy emitted in the B band is 37%, with lower and upper quartiles of 30% and 47%. This indicates that approximately 63% of the B band energy in these flares was in the continuum.

## 11. Temporal evolution of Balmer emission lines during flares

As mentioned in the introduction, stellar flares have been modelled as a combination of a short-lived rise in the continuum followed by a slower increase in hydrogen Balmer emission. Our typical spectral integration time of 300 seconds limits our ability to resolve events in time, as calculations of flux are quantified per spectrum. The smaller the time difference between events, the lower the probability they would occur during different spectra and thus be resolved. In less energetic flares where spectra have lower SNR, the sequence of events is also less clearly defined. To investigate temporal evolution during flares, we have therefore again focused on the three largest flares which all have B band flare energies greater than $10^{32}$ erg.

For each of these flares we calculated how the integrated flare flux in the Hα, Hβ, Hγ, Hδ, and He I 5876 emission lines changed as the flares progressed. We also calculated the changing flare flux level in the continuum adjacent to each line. Figure 8 shows how the flare flux in these emission lines and in the adjacent continuum varied as a function of time since each flare started. Line flux in each spectrum is marked as connected dots in red, continuum flux similarly in blue.

We described earlier how we estimated uncertainty in the spectral flux at Balmer emission lines from the standard deviation of flux in our quiescent spectra and from the uncertainty in defining the interpolated continuum under these lines. By combining these flux uncertainties in our flare and quiescent spectra, we calculated uncertainties in our flare-only spectra for the flux in Balmer emission lines and in the continuum flux at these lines. In Figure 8, one standard deviation of uncertainty in line flux is shown as red bands and in continuum flux as blue bands. In general, uncertainties increase as the flux in spectra decreases. Although, as we shall see in Figure 10, growth in the continuum at Hα in flares tended to be small, the Hα emission lines in these flares grew strongly and could be well measured, as shown in Table 5.

During the largest flare on 21 November 2021, each of the emission lines peaked one spectrum later than their adjacent continuum. In the other two flares, emission lines and continuum peaked during the same spectrum. There were two flares on 26 November 2021 (see Figure 1), with flux dropping to almost zero between them. It is notable that flux in the continuum decayed more quickly following the peak than flux in the Balmer lines. It appears in Figure 8 that there is a pattern with shorter wavelength Balmer lines decaying more quickly. Flux in the He I 5876 Å line remained high for longer than the Balmer lines before then decaying rapidly. This is similar to behavior reported in Hawley and Pettersen (1991) for AD Leo. We also noted that the peak in B band photometry always occurred during the same spectrum as the peak in continuum flux. This may be expected as the peak in continuum flux is a major driver for the photometric peak.

To quantify the tendency for shorter wavelength Balmer lines to decay more quickly, we measured the $t_{1/2}$ times of the Hα to Hδ Balmer lines in the three largest flares. This is the time interval between half maximum flux on the line rising and the same height on its decay, in other words the duration of the line measured at half maximum. These times are listed in Table 6. Uncertainties in flux are propagated into uncertainties in time. Figure 9, which plots these times along with linear fits to the data for each flare, shows that the duration of flares in Balmer lines is indeed positively correlated with their wavelength. This behavior is similar to that shown in Figure 18 and related text in Kowalski *et al.* (2013). Note also that the $t_{1/2}$ times of the Balmer lines in flares are several times longer than the $t_{1/2}$ times measured in the peaks of B band photometry given in Table 3. Again, this is consistent with the continuum decaying faster during flares relative to the decay in Balmer emission.

## 12. Spectral evolution of Balmer emission lines

Figure 10 compares Balmer line profiles in flux calibrated spectra at flare peak and quiescence on 21 November 2021. This shows that absolute flux in the Hβ, Hγ, and Hδ lines and in the continuum adjacent to these lines increased considerably relative to the quiescent level during the flare, whereas at Hα the continuum in quiescence was already higher and increased relatively little during the flare. Flare energy in the Hα line was broadly similar to that in the Hβ line as Table 5 shows.

To measure the Full Width at Half Maximum (FWHM) in Angstroms of Balmer emission lines during flares, we fitted Gaussian profiles to the Balmer lines in flare-only spectra after subtracting the interpolated continuum under the line. In most cases the line profiles were well fitted by a Gaussian profile, but in spectra at the peak of the larger flares we saw an excess of flux in the wings of the lines, particularly towards shorter wavelengths. In Figure 11 we show Gaussian fits to the Hβ line in the first three spectra of the largest flare peak on 21 November 2021. In the first two spectra there is clearly additional emission in the form of low wings which are more extensive on the blue side of the line and reach to around –1500 km/s. Although these wings are relatively poorly defined in our spectra, we attempted to model them by including an additional wide, low amplitude Gaussian component in the fits for the first two spectra. We found that the peaks of these additional components were displaced by around –100 km/s relative to the Hβ line and had FWHM of ~1600 km/s. This suggests that there was short-lived, blue-shifted emission in the Hβ line at the start of the flare.

In Figure 12 we show how FWHM of the Hα to Hδ lines varied during the course of the large flare on 21 November 2021. After a brief expansion, the lines rapidly settled back to their pre-outburst width.

To investigate the relationship in time between the changing flux (in Figure 8) and width (in Figure 12) of the Balmer lines as a flare evolves, we show in Figure 13 plots of flux vs FWHM for the Hα to Hδ lines during the peak of the large flare on 21 November 2021. The trajectories all follow a counter-clockwise loop whose direction of travel is marked with an arrow. All lines except Hα reach their maximum width in the spectrum before the lines reach their peak flux.

## 13. Summary and conclusions

Working as a collaborative group of small telescope scientists, we observed 12 flares of the dwarf M star EV Lac

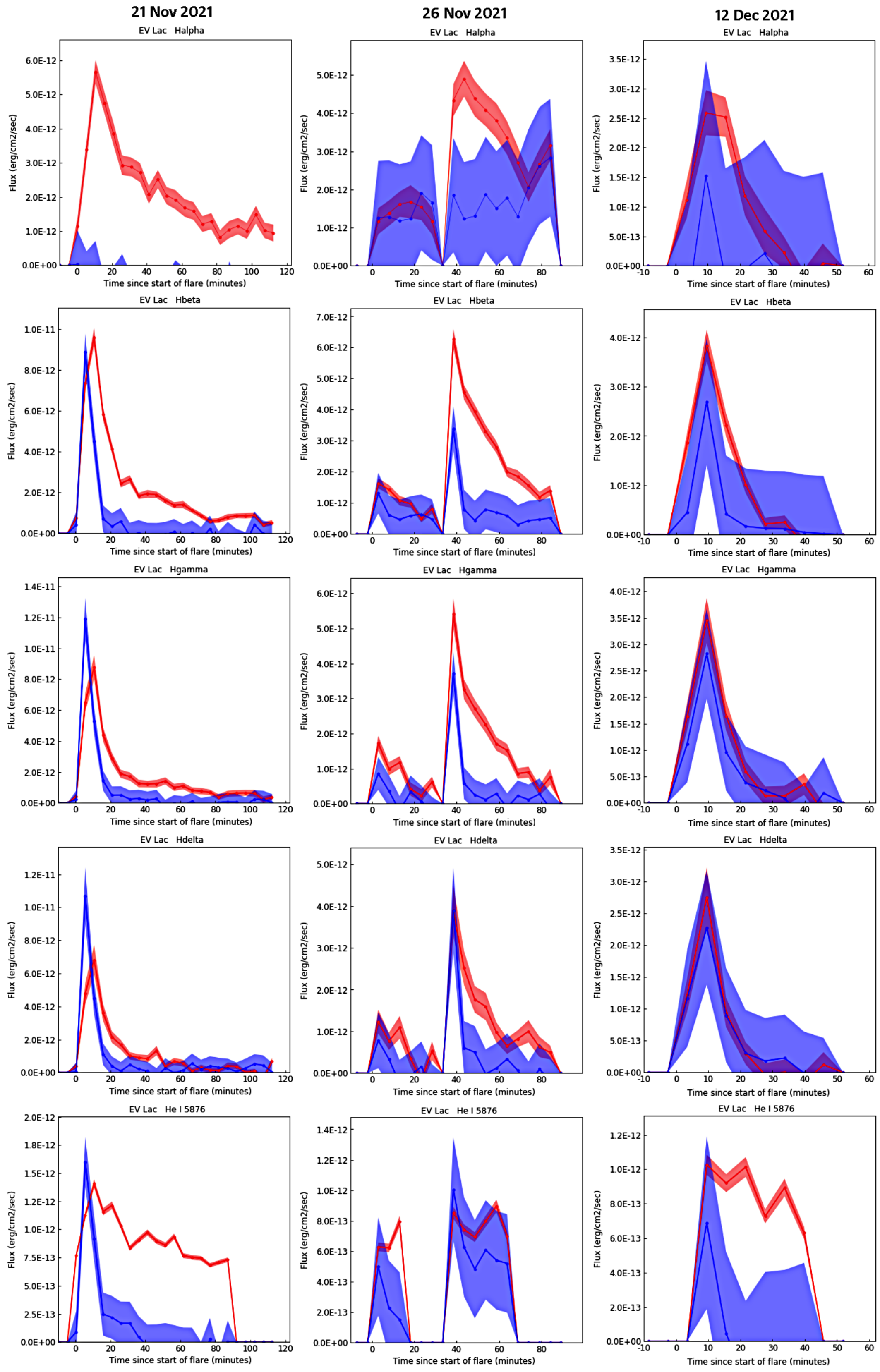

Figure 8. Temporal evolution of emission line flux and continuum flux in the three largest flares.

Table 6. $t_{1/2}$ times of Balmer emission lines in the three largest flares.

| *Flare No,* | *Date* | *Hα $t_{1/2}$ (min)* | *Hβ $t_{1/2}$ (min)* | *Hγ $t_{1/2}$ (min)* | *Hδ $t_{1/2}$ (min)* |
|---|---|---|---|---|---|
| 7 | 2021-Nov-21 | 32 ± 3 | 16 ± 2 | 12 ± 2 | 13 ± 2 |
| 9 | 2021-Nov-26 | 36 ± 5 | 20 ± 4 | 13 ± 4 | 12 ± 2 |
| 10 | 2021-Dec-12 | 18 ± 2 | 15 ± 2 | 12 ± 2 | 10 ± 2 |

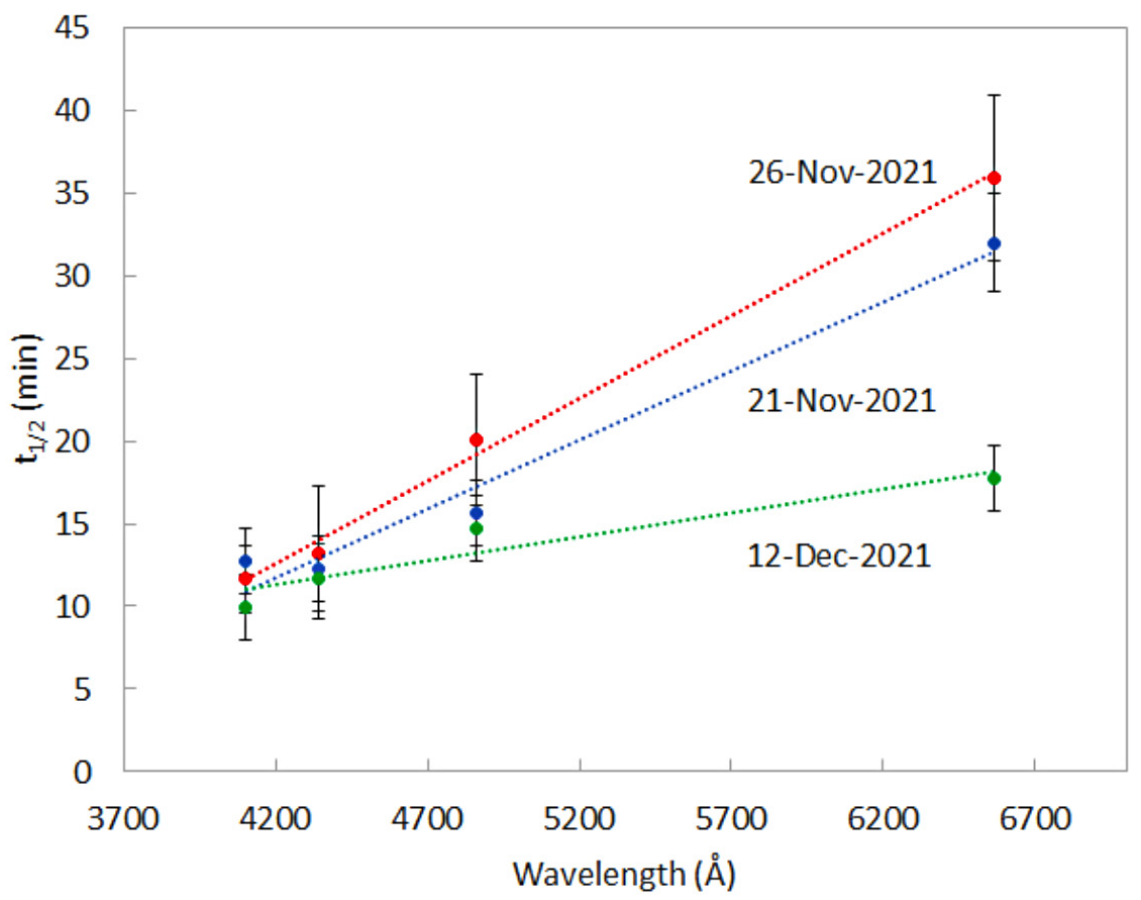

Figure 9. $t_{1/2}$ time vs wavelength for the three largest flares showing a positive correlation between Balmer line flare duration and wavelength.

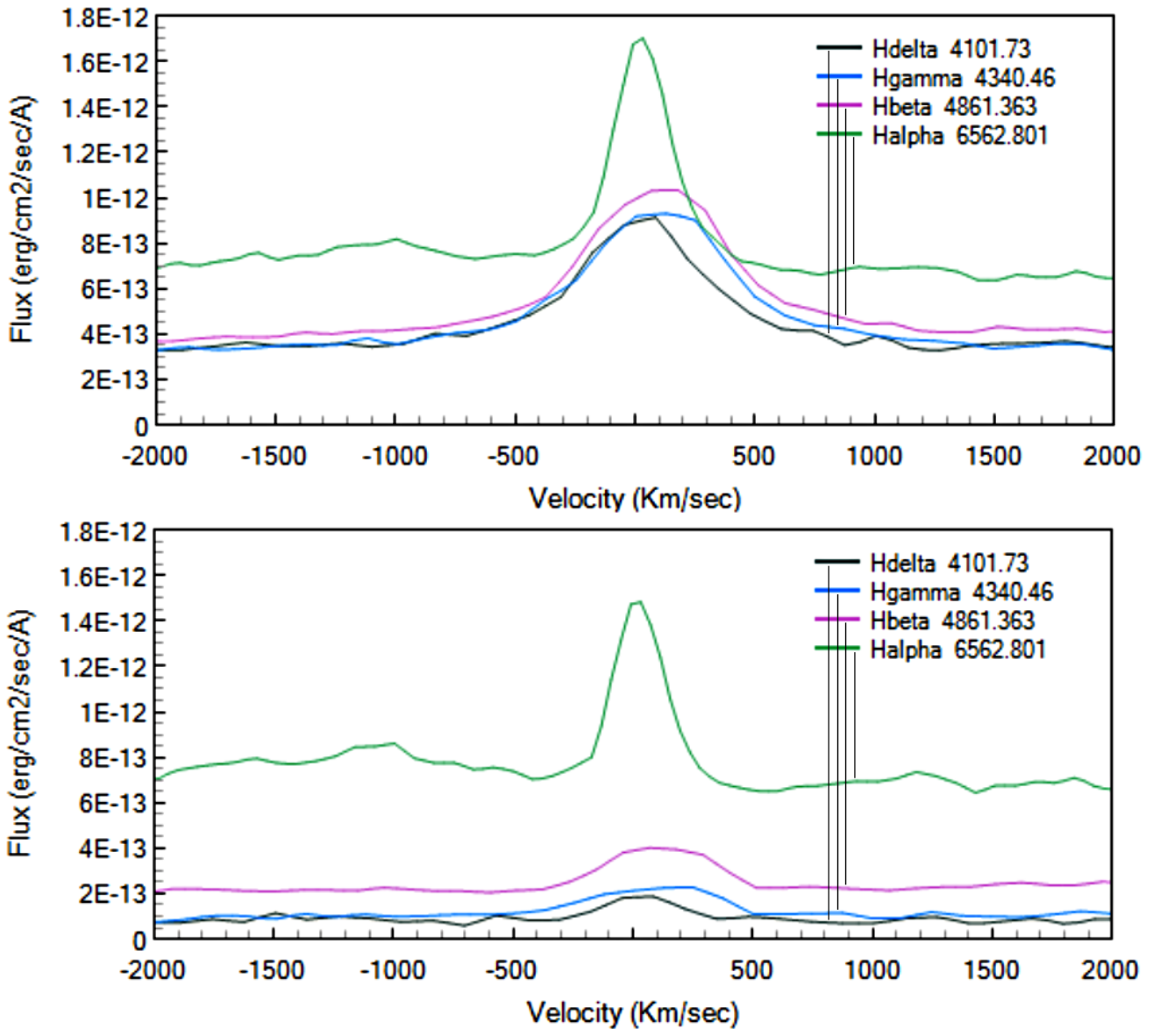

Figure 10. Balmer line profiles at flare peak (upper) and quiescence (lower) on 21 November 2021.

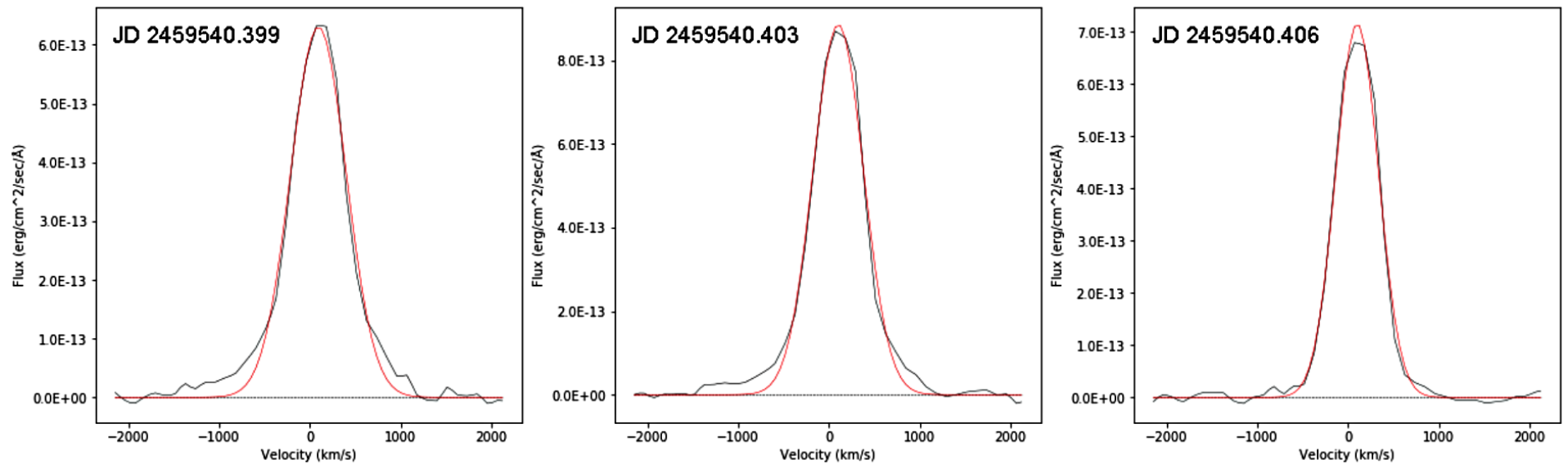

Figure 11. Gaussian fits to Hβ emission lines in the first three flare-only spectra during the flare peak on 21 November 2021. Data are marked as a solid black line, the Gaussian fit as a solid red line, and the continuum level as a dotted black line.

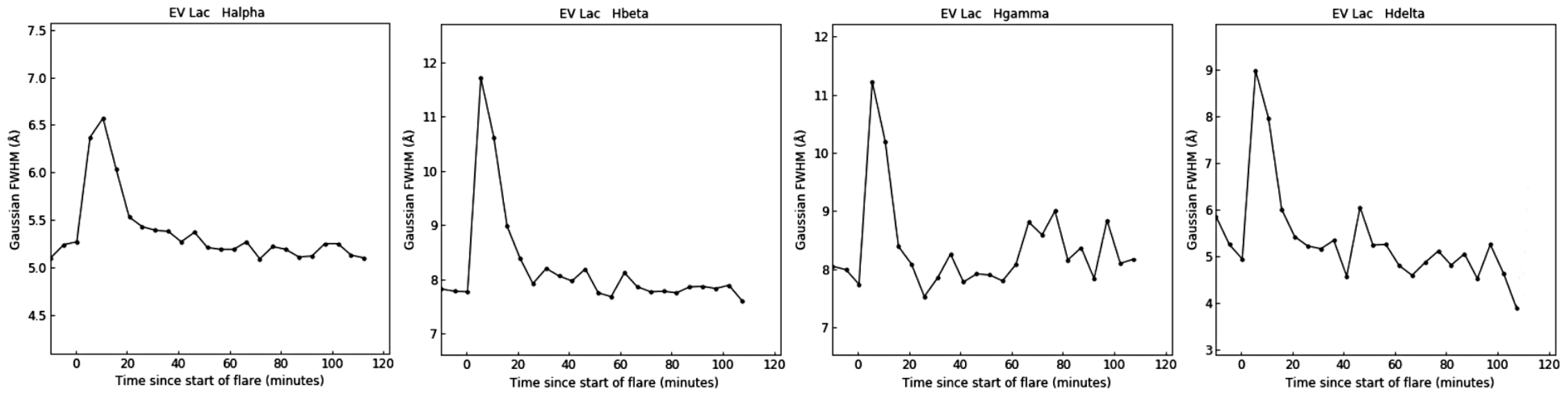

Figure 12. Evolution of FWHM in Hα to Hδ lines during the flare on 21 November 2021.

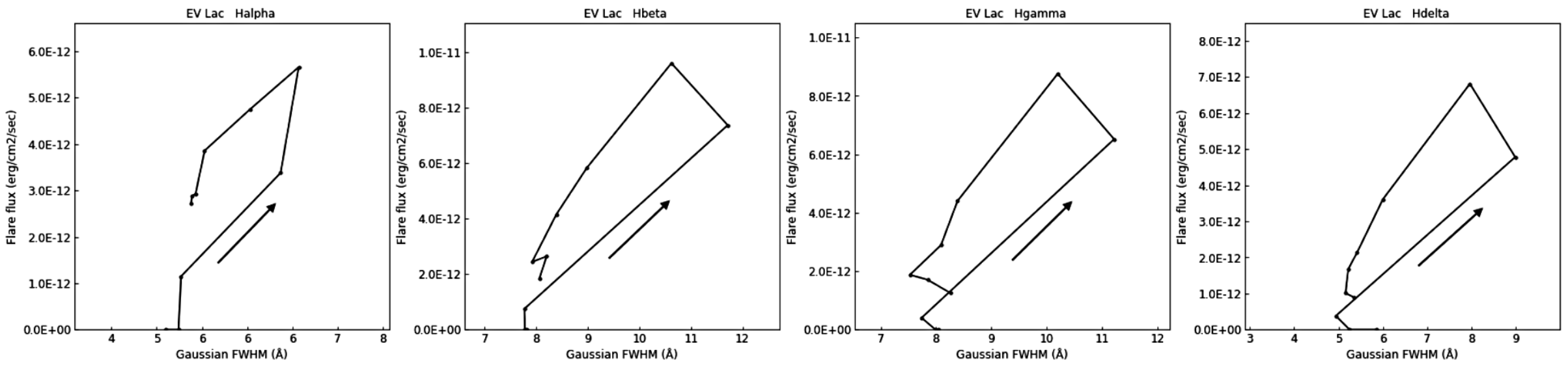

Figure 13. Relationship between flare flux and FWHM for the Hα to Hδ lines during the flare peak on 21 November 2021. The arrows show the direction of travel in time.

with B-band amplitude greater than 0.1 magnitude for which we concurrently recorded low resolution spectroscopy and B-band photometry. We calibrated our spectra in absolute flux using the B-band photometry and calculated B-band flare energies in the range Log E = 30.8 to 32.6 erg. We subtracted mean quiescent spectra to obtain flare-only spectra, calculated the energy emitted in Balmer emission lines during each flare, and monitored how this changed as flares evolved. Although our time resolution was limited by the length of our spectral exposures (300 sec), we observed in the brightest flare that flux in the continuum clearly peaked before flux in the Balmer emission lines. We found that flux in the continuum decayed faster than flux in emission lines and that shorter wavelength Balmer lines decayed faster. By fitting a Planck function to the blue continuum of the three brightest flares, we obtained their black-body temperatures.

Several publications (for example Alekseev *et al.* 1994; Abdul-Aziz *et al.* 1995; Osten *et al.* 2005; Paudel *et al.* 2021) have reported on optical band photometric and spectroscopic observations of EV Lac. These have mostly used meter-class telescopes and have rarely managed to obtain concurrent photometric and spectroscopic observations because of constraints on observing schedules. We have attempted to remedy that deficit through a coordinated campaign of concurrent photometric and spectroscopic observations of flares using amateur-sized telescopes. Our data can be compared with and potentially used to constrain the predications of flare models.

## 14. Acknowledgements

We thank our referee for a very thorough and helpful review. We acknowledge with thanks the resources provided by the AAVSO and BAA in support of the amateur community. Our research made use of NASA's Astrophysics Data System Bibliographic Services and the SIMBAD database operated at CDS, Strasbourg, France. The work of the Astropy collaboration has been extensively used in the analysis of our data.